\documentclass[conference]{IEEEtran}

\usepackage{amsmath,amsxtra,amssymb}
\usepackage{epsfig,graphicx,cite,url,psfrag}
\usepackage[small,bf,up]{caption}
\ifCLASSINFOpdf
\else
\fi

\begin{document}

%\title{Bare Demo of IEEEtran.cls for Conferences}
\title{VANET Connectivity Analysis}
%\author{M. Kafsi, O. Dousse, P. Papadimitratos, T. Alpcan and J-P. Hubaux}

%\author{\IEEEauthorblockN{Michael Shell}
%\IEEEauthorblockA{School of Electrical and\\Computer Engineering\\
%Georgia Institute of Technology\\
%Atlanta, Georgia 30332--0250\\
%Email: http://www.michaelshell.org/contact.html}
%\and
%\IEEEauthorblockN{Homer Simpson}
%\IEEEauthorblockA{Twentieth Century Fox\\
%\and
%\IEEEauthorblockN{James Kirk\\ and Montgomery Scott}
%\IEEEauthorblockA{Starfleet Academy\\
%San Francisco, California 96678-2391\\


\author{\IEEEauthorblockN{Mohamed Kafsi\IEEEauthorrefmark{1}, Panos Papadimitratos\IEEEauthorrefmark{1},
Olivier Dousse\IEEEauthorrefmark{2}, Tansu Alpcan\IEEEauthorrefmark{3}, Jean-Pierre Hubaux\IEEEauthorrefmark{1}}
\IEEEauthorblockA{\IEEEauthorrefmark{1} EPFL, Lausanne, Switzerland}
\IEEEauthorblockA{\IEEEauthorrefmark{2} Nokia Research Center, Lausanne, Switzerland}
\IEEEauthorblockA{\IEEEauthorrefmark{3}T-Labs, Berlin, Germany}
}

%% Propositions, theorems, lemmas, etc-------
\newtheorem{prop}{Proposition}
\newtheorem{property}{Property}
\newtheorem{lemma}{Lemma}
\newtheorem{theorem}{Theorem}
\newtheorem{definition}{Definition}
%% MATH -----------------------------------------------------------
\def\Real{\mathbb{R}}
\def\Ints{\mathbb{Z}}
\def\Nats{\mathbb{N}}
\def\Comps{\mathbb{C}}
\def\E{\mathbb{E}}
\def\P{\mathbb{P}}

\newcommand{\dis}{\displaystyle}
\newcommand{\ie}{{\em i.e., }}
\newcommand{\eg}{{\em e.g., }}
\newcommand{\cqfd}{\hfill $\Box$}
\newcommand{\la}{\lambda}
\newcommand{\calr}{\cal R}
\newcommand{\calA}{\mathcal A}
\newcommand{\calB}{\mathcal B}
\newcommand{\calC}{\mathcal C}
\newcommand{\calD}{\mathcal D}
\newcommand{\calE}{\mathcal E}
\newcommand{\calF}{\mathcal F}
\newcommand{\calG}{\mathcal G}
\newcommand{\calI}{\mathcal I}
\newcommand{\calN}{\mathcal N}
\newcommand{\calS}{\mathcal S}
\newcommand{\calU}{\mathcal U}
\newcommand{\calV}{\mathcal V}
\newcommand{\calW}{\mathcal W}

\newcommand{\vx}{\vec{x}}
\newcommand{\vy}{\vec{y}}
\newcommand{\vz}{\vec{z}}
\newcommand{\va}{\vec{a}}

\newcommand{\mif}{\textrm{ if }}
\newcommand{\mthen} {\textrm{ then }}
\newcommand{\melse} {\textrm{ else }}
\newcommand{\mand} {\textrm{ and }}
\newcommand{\mor} {\textrm{ or }}
\newcommand{\mst} {\textrm{ such that }}
\newcommand{\mfa} {\textrm{ for all }}
%% =================================
\renewcommand{\captionfont}{\small}
%\addtolength{\abovecaptionskip}{-5mm}
%\addtolength{\intextsep}{-5mm}

% Alter some LaTeX defaults for better treatment of figures:
    % See p.105 of "TeX Unbound" for suggested values.
    % See pp. 199-200 of Lamport's "LaTeX" book for details.
    %   General parameters, for ALL pages:
    \renewcommand{\topfraction}{0.9}    % max fraction of floats at top
    \renewcommand{\bottomfraction}{0.8} % max fraction of floats at bottom
    %   Parameters for TEXT pages (not float pages):
    \setcounter{topnumber}{2}
    \setcounter{bottomnumber}{2}
    \setcounter{totalnumber}{4}     % 2 may work better
    \setcounter{dbltopnumber}{2}    % for 2-column pages
    \renewcommand{\dbltopfraction}{0.9} % fit big float above 2-col. text
    \renewcommand{\textfraction}{0.07}  % allow minimal text w. figs
    %   Parameters for FLOAT pages (not text pages):
    \renewcommand{\floatpagefraction}{0.7}  % require fuller float pages
    % N.B.: floatpagefraction MUST be less than topfraction !!
    \renewcommand{\dblfloatpagefraction}{0.7}   % require fuller float pages
    % remember to use [htp] or [htpb] for placement

\maketitle

\begin{abstract}
Vehicular Ad Hoc Networks (VANETs) are a peculiar subclass of mobile ad hoc networks that raise a number of technical challenges, notably from the point of view of their mobility models. In this paper, we provide a thorough analysis of the connectivity of such networks by leveraging on well-known results of percolation theory. By means of simulations, we study the influence of a number of parameters, including vehicle density, proportion of equipped vehicles, and radio communication range. We also study the influence of traffic lights and roadside units. Our results provide insights on the behavior of connectivity. We believe this paper to be a valuable framework to assess the feasibility and performance of future applications relying on vehicular connectivity in urban scenarios.
\end{abstract}


\section{Introduction}

Vehicular Ad Hoc Networks (VANETs) are a type of mobile ad hoc networks, with mobile nodes being vehicles and Road Side Units (RSUs) as static nodes. All car makers are currently investigating the feasibility and benefits of VANETs. Several prototypes have already been implemented and deployed, but there is consensus on the fact that VANETs still raise formidable design challenges. Predicting the level of VANET connectivity is a notable one, as it is strongly influenced by the peculiarities of VANETs, including the large range of possible node speeds, the mobility constrained by the road network, and the possible presence of traffic lights and RSUs.

\begin{figure}[htpb]
\begin{center}
\includegraphics[width=0.75\linewidth]{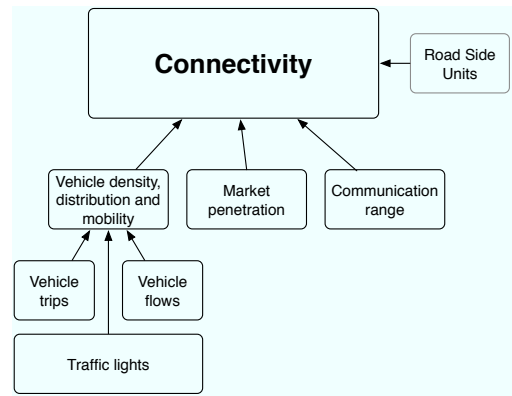}
\caption{Connectivity study approach.}
\label{fig:Research structure}
\end{center}
\end{figure}

In this paper, we study the impact of these unique characteristics on the connectivity of VANETs. More specifically, we assess the connectivity in
(two-dimensional) urban scenarios, as one-dimensional, simpler highway scenarios were already investigated. By means of well-known results of
percolation theory, we compute the probability for two nodes to be in power range of each other. We then derive the proportion of vehicles in the biggest cluster of the connectivity graph. We corroborate and extend this analysis by a carefully chosen set of simulations, in which we vary the parameters of vehicle density, market penetration of the communication equipment, and radio communication range, as well as several possible combinations of vehicle trips and vehicle flows. We finally study the impact of RSUs and of traffic lights. The overall approach is summarized in Fig.~\ref{fig:Research structure}.

One of the difficulties with VANETs is that it is  nowadays still unclear which applications will prevail, although it is usually admitted that collision avoidance and traffic optimization will be among the most prominent ones. To circumvent this problem, we take a rather \emph{agnostic} approach, assuming it is desirable for vehicles to communicate with each other and with RSUs (in a multi-hop fashion whenever needed), without worrying about the purpose of that communication. We also do not distinguish between unicast, multicast, broadcast, or geocast communications.

The rest of the paper is organized as follows. Sec.~\ref{Sec:RelWork} presents the related work. Sec.~\ref{Sec:Models} introduces the different models. Sec.~\ref{Sec:EvalConneAd-hocNet} relates the percolation theory to the connectivity in a square lattice and analyzes the connectivity results in a
pure ad-hoc (vehicle-to-vehicle) network. We introduce RSUs in Sec.~\ref{Sec:ConnectivityRSU} and investigate their effect on connectivity. Finally, Sec.~\ref{Sec:conclusion} summarizes the lessons learned from this investigation and concludes the paper.

\section{Related Work}\label{Sec:RelWork}

Connectivity in VANETs has been a topic of interest, especially due to the recently increasing research activity. Artimy et al. investigated connectivity in VANETs for both highway and simple road configurations, using the \textit{Roadsim} traffic micro-simulator to generate vehicle movement in multi-lane and unidirectional highways based on a cellular automaton model \cite{zpr92-115}. They examined factors  such as vehicle density, relative velocity, and number of lanes, and their influence on connectivity and in particular the maintenance of an active communication session between a pair of vehicles~\cite{ConnectivityinINterVehicle}. Subsequently, in \cite{Connectivitywithstatictransmissionrange}, they further investigated the effect of vehicle density and road configuration on the value of minimum transmission range, showing that an increase of vehicle density may not always result in higher connectivity. In a related study, an algorithm was proposed to adapt dynamically the vehicle transmission range based on local density estimation \cite{1080761}.

In \cite{VehiclesPortland}, Marfia et al. considered a network where all vehicles opportunistically exploit open access points to communicate with
other vehicles, looking at how the infrastructure can help inter-vehicular communication and how the stop-and-go behavior of vehicles can
increase network congestion and eventually lead to collapse in performance. They observed that a significant impact of the the mobility model on performance and
significant differences between real data and traffic simulators. In another study~\cite{4300800}, mobility scenarios in urban
settings are studied and the importance of accurate vehicle distribution modeling on network performance is identified. Ho et al. have studied in~\cite{1306785} the
connectivity of a vehicular ad-hoc network consisting of buses, using a hybrid simulator \textit{GrooveNet}~\cite{4205298} developed by Carnegie Mellon
University and General Motors, considering the impact of the topology, traffic signals, and vehicle traffic on network connectivity. In a more recent work \cite{jeromemobihoc08}, five analytical mobility models for vehicular networks, developed and used earlier in the literature, are investigated: through simulations, it is assessed how realistic the traffic they generate is and what the resultant connectivity graph properties are.

The connectivity analysis of VANETs is closely related to investigations on the connectivity of ad-hoc and hybrid networks. For example, Bai et al. proposed a framework for analyzing the impact of mobility on performance of routing protocols for ad-hoc networks metrics~\cite{DBLP:conf/infocom/BaiSH03}. Dousse et al. \cite{dousse02connectivity} used percolation theory to develop models capturing the behavior of connected clusters as a function of node density. In a related study ~\cite{PhysRevLett.85.4626,PhysRevLett.86.3682}, Cohen et al. investigated the resilience of the Internet to breakdowns and intentional attacks, using percolation theory to derive a lower bound on deactivated nodes that render the network disconnected.
\section{Models} \label{Sec:Models}

\subsection{Traffic model}

\begin{figure}[t]
\begin{center}
\includegraphics[width=0.75\linewidth]{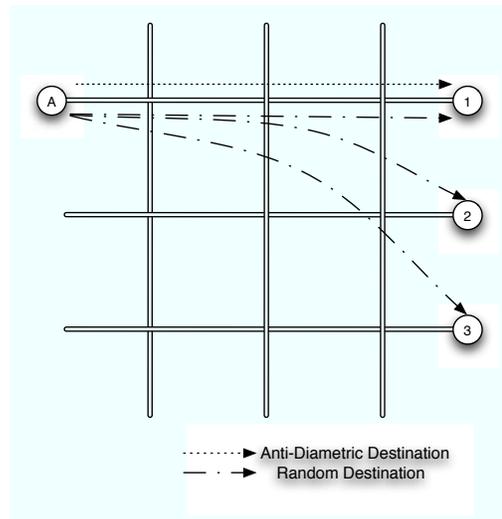}
\caption{The simulation domain for $N=3$. In the random destination scenario, vehicles entering at Point~A pick a destination uniformly between Points~A, B and C.}
\label{fig:domain}
\end{center}
\end{figure}

We consider a hash-shaped grid of $N$ vertical and $N$ horizontal roads with $N^2$ intersections, as depicted in Fig.~\ref{fig:domain}. The
distance between two consecutive horizontal (similarly, vertical) roads is $L$ meters. We consider three types of intersections:
\begin{enumerate}
\item \textbf{No Traffic Lights:} Vehicles yield to any vehicle arriving from their
right.
\item \textbf{Synchronized Traffic Lights:} Intersections are equipped
with traffic lights. All intersections are synchronized, so that the light
is green in the horizontal direction at all intersections at the same time.
\item \textbf{Alternating Traffic Lights (``Green Wave''):}
Intersections are equipped with traffic lights. We adjust the traffic light phase duration
to simulate the green wave effect, so that drivers can go through successive intersections in a main direction without stopping.
\end{enumerate}

%\footnote{This is the default rule in Europe. Note that it
%slightly differs from the ``four way stop'' scheme used in the United States,
%as vehicles do not necessarily traverse the intersection in the same order as they
%arrive.}

In our model, vehicles enter the simulation domain with rate $f$ from all $4N$ road end points. The number of horizontal and vertical
perpendicular roads range in our investigation from $N=5$ to $10$. Vehicles enter the simulation domain at times chosen uniformly throughout the simulation duration. If the simulation time is sufficiently long, vehicle arrivals can locally seen as a Poisson process of intensity $f$. Each entering vehicle picks its destination (exit point) in two different ways:
\begin{enumerate}
\item \textbf{Anti-Diametric Destination:} The vehicle source (entry) and destination (exit point) are at the
same row or column of the road network.
\item \textbf{Random Destination:} Vehicles select uniformly their destination among the $N$ road end points
at the opposite side of the road network.
\end{enumerate}
After reaching their destination, the vehicles leave the simulation domain. The total number of vehicles at time $t$ is denoted by $n(t)$.

We use SUMO~\cite{sumo}, as microscopic, continuous-space and discrete-time traffic simulator to model realistic vehicle behavior. SUMO simulates the movement of every single vehicle implementing \emph{vehicle following model}. The driver behavior depends on the behavior and the speed of the preceding car and a \emph{dynamic user assignment} to find out the route to take. The driver model decides in real time which route to take and her choice is not only dictated by the route length but also by the average speed of other vehicles along the same route. This leads to realistic behavior, as drivers change routes to avoid congestion, preferring, for example a longer but less congested route. Interestingly, as our experiments reveal, for anti-diametric destinations, chosen routes do not include turns in most cases, because the vehicle traffic is mostly uniform across the road network.
\begin{table}[t]
\begin{center}
\begin{tabular}{|l|l|}
\hline
Name & Notation [Unit]\\
\hline
\hline
Vehicle arrival rate & $f [h^{-1}]$\\
\hline
Market penetration & $\rho[-]$\\
\hline
Connectivity range & $r [m]$\\
\hline
Road segment length & $L [m]$\\
\hline
Isolated vehicles (\%) & $ \phi [-]$\\
\hline
Vehicles in largest component (\%) & $ \theta [-]$\\
\hline
\end{tabular}
\caption{Key model parameters.}\label{table:param}
\end{center}
\end{table}
In the \emph{stationary regime}, i.e., long after the beginning of the simulation), the rate at which vehicles leave the domain is equal to $f$ after stabilization of the simulation, unless the throughput of vehicles cannot be supported by the roads. In this case, \emph{saturation} occurs, i.e., entry road segments fill up with vehicles and the simulator fails to add new vehicles. Finally, we introduce $0< \rho \le 1$, the fraction of equipped vehicles (essentially market penetration for vehicular communications). The average number of equipped vehicles is $\rho n(t)$.

We present results for the three types of road traffic control differentiated by the existence and operation of traffic lights: ``\textbf{N}o Traffic \textbf{L}ights'' (\textbf{NL}), ``\textbf{S}ynchronized Traffic \textbf{L}ights'' (\textbf{SL}), and ``\textbf{G}reen \textbf{W}ave'' (\textbf{GW}). We combine those settings with different road destination selection settings: ``\textbf{A}nti-diametric'' or ``\textbf{R}andom'' destination road end points. For example, SL-R signifies synchronized traffic lights with random destinations. Note that the green wave regime can be meaningful only when vehicles can keep a constant speed. This is not the case for the random destination model, where vehicles are more likely to turn at intersections, decelerate or even stop before doing so, and thus cause significant velocity fluctuations; this is why we do not investigate the GW-R case.

\subsection{Connectivity Model}

We assume that two vehicles communicate directly through a wireless link if their distance does not exceed a certain \emph{connectivity range} $r$. As a result, we obtain at each point in time a connectivity graph, the main object of this study, whose vertices correspond to vehicles equipped with an on-board vehicular communication unit. We use below the terms ``vehicle'' and ``nodes'' interchangeably, unless noted otherwise. The connectedness of the graph is measured with two (observed) values: (i) the fraction of (equipped) vehicles that are not connected with any other vehicle, denoted by $\phi(t)$, and (ii) the fraction of vehicles that belong to the largest connected component of the graph, $\theta(t)$. The parameters of our model are summarized in Table~\ref{table:param}.

\section{Vehicle-to-Vehicle Connectivity} \label{Sec:EvalConneAd-hocNet}

This section presents our findings on vehicle-to-vehicle (V2V) connectivity, based on analysis and extensive experiments with microscopic vehicle mobility simulation. At first, we seek to understand how vehicles are distributed across the road grid, as a function of their entrance rate. Our findings, detailed in Sec.~\ref{sec:CarDensityfctCarFlow}, show that traffic has a significant impact on the density and spatial distribution of vehicles. The differing vehicle placement conditions, as a result of the differing rates of entrance (and exit) from the simulated road grid, influence the overall connectivity and most notably the proportion of vehicles in the largest cluster. But, as it will be explained in Sec.~\ref{sec:cluster}, traffic lights significantly influence the size of largest cluster. On a different dimension of our investigation, we consider the \emph{critical range}, that is, the minimum communication range that enables all vehicles to be part of a single cluster. As shown in Sec.~\ref{sec:critical_range}, traffic lights heavily influence the value of the critical range. We shift gears in Sec.~\ref{sec:market_penetration}, where we are concerned with the presence of vehicles that are not equipped with radios but still affect the spatial distribution of the equipped vehicles and thus the vehicular network connectivity. The simulation parameters are summarized in Table~\ref{table:simparam}.

\begin{figure*}[t]
\includegraphics[width=\linewidth]{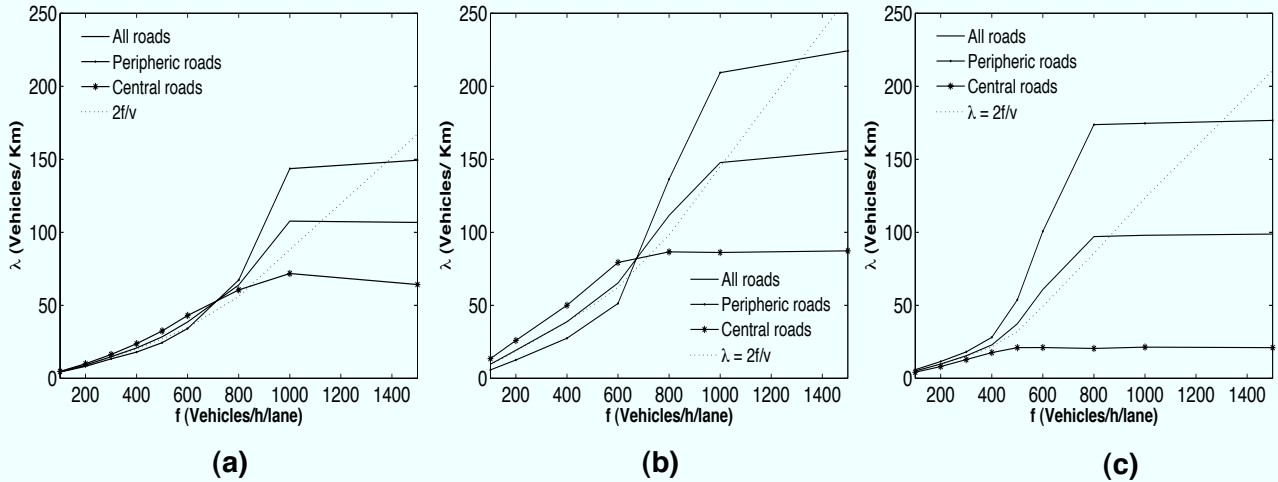}
\caption{ Vehicle density vs vehicle flow in the scenarios (a) NL-A (b) SL-A and (c) GW-A.} \label{fig:Car_density_Scenario1&2&3}
\end{figure*}

\begin{table}[t]
\begin{center}
\begin{tabular}{|l|l|}
\hline
Simulation interval & [0,999s] \\
\hline
Number of lane & 2  \\
\hline
Grid dimension & 5, 10 \\
\hline
Traffic lights regimes & Synchronized, Green Wave\\
\hline
Road segment length $L$ [m] & 400 \\
\hline
$f$ (Vehicles/h/lane) & 100..1500\\
\hline
Types of destination & Anti-Diametric, Random\\
\hline \hline
Connectivity range $r$ [m] & 100\\
\hline
\end{tabular}
\caption{Simulation parameters}\label{table:simparam}
\small\addtolength{\tabcolsep}{-5pt}
\end{center}
\end{table}

\subsection{Vehicle Density} \label{sec:CarDensityfctCarFlow}

As the simulator allows us to control the in-flow rate for each source (entry) point, we utilize this as the independent parameter in our experiments.
But it is the resultant placement of vehicles that is of real interest. We investigate how the entrance rate (for a given $N$ dimension of
the road network grid) relates to the average vehicle density. We note nonetheless that vehicle density and spatial distribution more generally do not depend only on the flow of vehicles entering the network but also on the mobility regime and type of vehicle trips. We differentiate \emph{peripheral road segments} as those $4N$ segments between the vehicle entry (exit) points and the first- (last-) encountered intersection. We term all the remaining road segments in the grid as \emph{central road segments}. This is motivated by the observation that, as explained below, there is a variation of the distribution across the road network.

We first look at the scenarios with anti-diametric destinations in Fig.~\ref{fig:Car_density_Scenario1&2&3}: as the vehicle in-flow rate increases, the vehicle density increases in a similar manner for peripheral and central road segments. But this is so only up to a certain value of $f$, e.g., $f=800 \,$vehicles/h/lane (in Fig.~\ref{fig:Car_density_Scenario1&2&3}.(a)), after which the density in central roads increases very slowly and then it does not increase any further. While, the vehicle density in peripheral roads continues increasing until it also reaches a plateau for a larger in-flow rate, e.g., $f=1000\,$vehicles/h/lane for Fig.~\ref{fig:Car_density_Scenario1&2&3}.(a). This phenomenon occurs consistently for all scenarios, but with different transition $f$ values. Basically, the central road segment $\lambda$ is lower for GW compared to SL and NL scenarios. At the same time, the differentiation between peripheral and central segments occurs for lower $f$ values for GW, compared to SL and NL.

These variations are due to the role of traffic lights. Looking, for example, at Fig.~\ref{fig:Car_density_Scenario1&2&3}.(c), central road density flattens at $f=500 \,$vehicles/h/lane: traffic lights at the first intersection encountered by vehicles, that is, at the limit of each peripheral road segment, control the in-flow to the central road segments. Essentially, they act as a bottleneck. At this point, traffic jams start building up at the peripheral roads, up to the point the density reaches, at $f=800 \,$vehicles/h/lane, a plateau: the saturated peripheral roads do not allow additional vehicles to the simulation domain. In contrast, without traffic lights, stop signs at intersections start acting as a bottleneck only when the amount of traffic is such that long queues build up at all sides of the intersection. This essentially ``shifts'' the saturation point for $\lambda$ to occur for $f>1000 \,$ vehicles/h/lane. But it is interesting to see that for NL, the plateau occurs for peripheral and central roads for that same $f$. Again, peripheral roads reach higher densities, up to the point they get saturated.

\begin{figure*}[t]
\begin{center}
\includegraphics[width=0.8\linewidth]{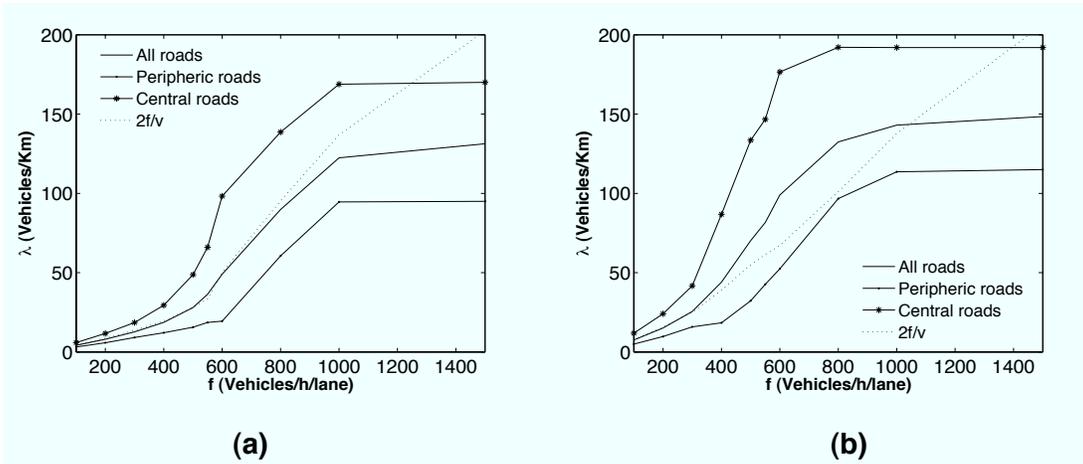}
\caption{Vehicle density vs. vehicle flow in the scenarios (a) NL-R, (b) SL-R.} \label{fig:Car_density_Scenario4&5}
\end{center}
\end{figure*}

For scenarios with random destination(s), in Fig.~\ref{fig:Car_density_Scenario4&5}, we also observe the saturation phenomenon and the differentiation between peripheral and central road segments. However, central roads get relatively more congested than peripheral roads. This peculiarity stems from the trip type: with random destinations vehicle routes are longer while vehicles are more likely to turn at intersections to reach their destination. As a consequence, vehicles stay longer in the (simulated) road network and especially in central road segments.

\subsection{Vehicle Clustering}\label{sec:cluster}

We study the clustering of vehicles both analytically and through simulations. For scenarios with simple intersections, we show that accurate predictions of the network connectivity can be made using percolation theory. For the scenarios involving traffic lights, we study connectivity primarily through simulations.

\subsubsection{Analytical study}
For low vehicle arrival rates, road traffic has little impact on the spatial distribution of vehicles. In the stationary regime, we can model this with a spatial Poisson process. We neglect the width of the streets and associate them with lines and we obtain a one-dimensional process along each road; its intensity,  $\lambda$, is related to the vehicle arrival rate through the formula:
$$
\lambda = \frac{2f}{\bar{v}},
$$
where $\bar{v}$ denotes the average speed of the vehicles and the factor $2$ is due to the bi-directionality of traffic.

\begin{figure}[t]
\begin{center}
\includegraphics[width=0.8\linewidth]{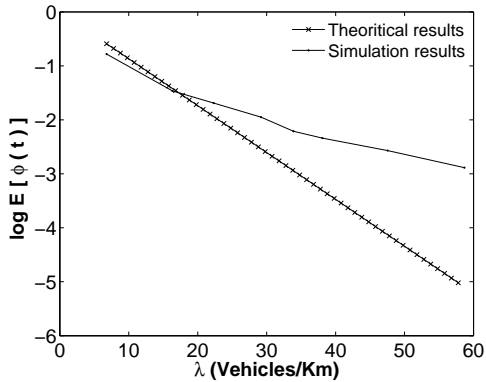}
\caption{Theoretical and simulated fraction of isolated vehicles in a scenario with no traffic lights and Anti-diametric destinations. $N=10$ and $L=400$m long.}
\label{fig:isolated}
\end{center}
\end{figure}

Given the Poisson assumption, one can compute an upper bound on the average fraction $E[\phi(t)]$ of vehicles that are connected to no other vehicles in the stationary regime, that is, when the road network is at a state that the rate of vehicles entering the road network is the same as the rate of vehicle leaving it:
$$
E[\phi(t)] = \exp(-2\lambda \rho r).
$$
Fig.~\ref{fig:isolated} compares the analytical curve with simulation results, showing the limitations of the Poisson approximation: When the density of vehicles is not very low, the intersections start shaping the vehicle spatial distribution. Vehicles that are stopped at an intersection are typically not isolated, whereas vehicles leaving an intersection are spaced apart and they are likely to be isolated.

To compute $\theta(t)$, the fraction of vehicles (nodes) belonging to the largest connected component, we start by looking at the probability $p$ that a road segment (i.e., the portion of road between two consecutive intersections) is covered by a sequence of connected vehicles. We can then see the network as an edge percolation model, where each road segment is \emph{covered} (or \emph{open}) with probability $p$.

Percolation theory allows us to address the case of a network with infinite size, with $N\rightarrow\infty$. Then it is well known that (see, for example, \cite{Grimmett1999}):
$$
E[\theta(t)] = 0 \mif p < \frac{1}{2}
$$
and
$$
E[\theta(t)] > 0 \mif p > \frac{1}{2},
$$
which is termed the \emph{percolation phenomenon}. In other words, there is a \emph{critical density} above which a giant connected component appears
(\emph{super-critical phase}). Below this density, all connected components have a finite size.

If the network is large but finite, its connectivity behavior is not much different. Below the critical density, all connected components are relatively small, and the largest of them may only contain a small fraction of the vehicles. On the contrary, above the threshold, a large connected component forms, containing typically more than half of the vehicles. Therefore, a ``soft'' transition is expected, with the fraction of vehicles in the largest cluster suddenly shifts from low values to values close to one.

We emphasize that these results are valid only if each and every road segment is covered independently with the probability $p$. In practice, however, the location of vehicles near intersections may influence the connectivity of adjacent segments. To cope with this dependence, we use two conditions, one necessary and one sufficient for connectivity. This provides us an upper and a lower bound on the connectivity respectively.

The necessary condition is obtained by making the optimistic assumption that there is at least one vehicle located at the center of each intersection. Thus the condition for the connectivity of a segment is to have a sequence of connected vehicles linking the two intersections. This depends only on the location of the vehicles on the segment itself and is thus independent of any other segment. We denote the probability this condition is
fulfilled for a given segment by $p_u$, which we compute with the help of Theorem~1 in~\cite{dousse02connectivity}:
$$
p_u = \left\{ \begin{array}{l}
1 \hspace{3.5cm} \mif 0 \leq L < r \\
\vspace{0.5ex}
\sum_{i=0}^{ \lfloor x/r \rfloor}
\frac{\left( -\lambda e^{-\lambda r} (x - ir) \right)^{i} }{i ! } \\
\;\;\; - e^{-\lambda r} \sum_{i=0}^{ \lfloor L/r \rfloor - 1}
\frac{\left( -\lambda e^{-\lambda r} (L - (i+1)r) \right)^{i} }{i ! } \\
\mbox{ } \hspace{3.8cm} \mif L \geq r \end{array} \right.
$$
The sufficient condition is obtained by assuming that each vehicle is the center of a disk of radius $r/2$ and requiring that the segment completely is covered by disks. Consecutive vehicles must be at least at distance at most $r$, to be connected. However, the ends of the segment might be covered by disks from adjacent segments, which implies the coverage also depends on the location of vehicles in neighboring segments. To avoid this dependence, we strengthen the condition, by requiring that the segment be covered by disks centered on the same segment. This leads to an independent condition that is sufficient for connectivity; it is fulfilled with probability $p_l$, which can also be computed from Theorem~1 in \cite{dousse02connectivity}:
\begin{eqnarray*}
p_l &=& \sum_{i=0}^{ \lfloor L/r \rfloor+1} \frac{\left( -\lambda
e^{-\lambda r} (L - (i-1)r) \right)^{i} }{i ! } \\
& & - e^{-\lambda r} \sum_{i=0}^{ \lfloor L/r \rfloor} \frac{\left( -\lambda
e^{-\lambda r} (L - ir) \right)^{i} }{i ! }.
\end{eqnarray*}

%%%%% Commented the two psfrag... add those to the graphic
%\begin{figure}[htpb]
%%\psfrag{L}{\footnotesize $L$}
%%\psfrag{r/2}{\footnotesize $r/2$}
%\begin{center}
%\includegraphics[width=0.8\linewidth]{balls.eps}
%\caption{Overlapping disks of radius $r/2$. If two adjacent segments are completely covered with balls, the vehicles at the center of each ball are connected.}
%\label{fig:balls}
%\end{center}
%\end{figure}

\begin{figure}[t]
\begin{center}
\includegraphics[width=0.8\linewidth]{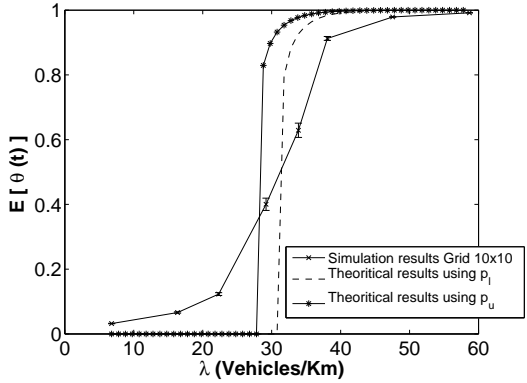}
\caption{Fraction of vehicles in the largest cluster; two theoretical curves and the simulation-based one, $N=10$ for the NL-A scenario.}
\label{fig:connectivity}
\end{center}
\end{figure}

Fig.~\ref{fig:connectivity} shows a comparison of the theoretical values of $E[\theta(t)]$, using the necessary and sufficient conditions, and simulation results. We see that despite the inaccuracy of the Poisson model in predicting detailed connectivity (isolated vehicles), global connectivity matches the model qualitatively. As expected, due to the finite size of the simulated network, the transition is much softer than the one in the theoretical model (for infinite networks). However, the threshold at which the fraction of vehicles in the largest cluster jumps is well
predicted by our model.

\begin{figure}[t]
\begin{center}
\includegraphics[width=0.8\linewidth]{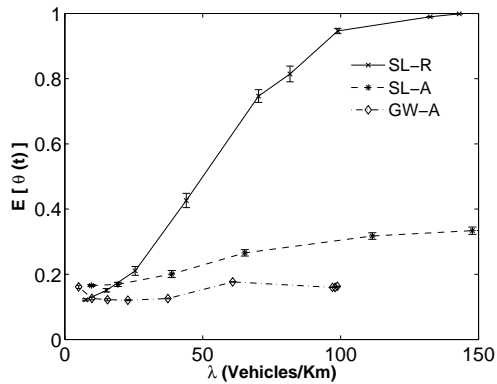}
\caption{Proportion of vehicles in the largest cluster vs. vehicle density.}
\label{fig:Biggest_cluster_function_car_density}
\end{center}
\end{figure}

\subsubsection{Simulation Analysis}

The theoretical analysis provided above hinges on the assumption that vehicles are uniformly distributed across the network area. But this is not the case for SL and GW, for the settings with traffic lights, and more so for the SL-R and GW-R scenarios. This is why we analyze the SL and GW cases only through simulations, both for anti-diametric and random destination scenarios. Fig.~\ref{fig:Biggest_cluster_function_car_density} shows the average proportion of vehicles in the largest cluster, as a function of vehicle density for SL and GW. We observe in Fig.~\ref{fig:connectivity} a major difference between SL-A and GW-A and NL-A. For NL-A, $E[\theta(t)]$ increases with density ($\lambda$), and reaches 1, with almost all vehicles connected. On the contrary, for traffic-light regulated transportation, that is, for GW-A and SL-A, $E[\theta(t)] < 0.35$.

%% #11# given the above statement, it would great here to have the SL-R and
%% GW-R

\begin{figure}[t]
\begin{center}
\includegraphics[width=0.8\linewidth]{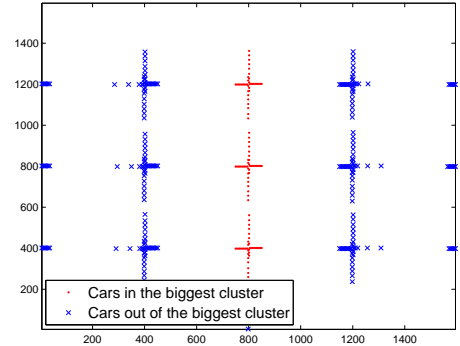}
\caption{Example of a typical biggest cluster in the scenario SL-A, corresponding to the ``release'' of the vehicles in
the vertical direction.}
\label{fig:Biggest_cluster_flow400_Grid_5_400_tl}
\end{center}
\end{figure}

This is explained if we observe the vehicle distribution and density for each scenario. Recall that for $\lambda < 60$ vehicles/Km, vehicles are equally distributed across central and peripheral roads. The network percolates when reaching the critical density: the vehicles form one giant cluster. For he scenarios with traffic lights, the largest cluster size is bounded: $0.2E[n(t)]$ and $0.35E[n(t)] $ for the SL-A and GW-A settings respectively. Interestingly, even though higher density, $\lambda$, is achieved (with increasing $f$ values) for SL-A than the density achieved for NL-A (as shown in Sec.~\ref{sec:CarDensityfctCarFlow}), the SL-A largest cluster size is lower than for NL-A. This can be explained by considering the traffic dynamics: For SL, there exist basically two traffic phases: traffic lights green, first, in the horizontal and then in the vertical direction. This results in the formation of lengthy horizontal and vertical disconnected clusters, as those shown in a snapshot of the network connectivity in Fig.~\ref{fig:Biggest_cluster_flow400_Grid_5_400_tl}. These ``parallel'' clusters of vehicles in motion never merge into a large and eventually giant cluster in the range of scenarios we investigate. Note that the network percolates in the SL-R scenario because the network reaches congestion, with the mean travel time being 25\% higher than that in SL-A). As the central road segments become really congested, vehicles traffic becomes almost static in the center of the domain. On the other hand, for the GW scenarios, the central road density is low and well under the critical density; in the sub-critical phase, connectivity remains poor as the network does \emph{not} percolate.

\begin{figure}[t]
\begin{center}
\includegraphics[width=0.8\linewidth]{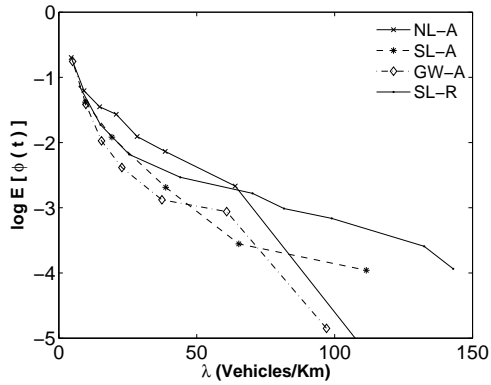}
\caption{Proportion of isolated vehicles vs. vehicle density.}
\label{fig:Isolated_cars_function_car_density}
\end{center}
\end{figure}

Fig.~\ref{fig:Isolated_cars_function_car_density} shows the logarithm of the average proportion of isolated vehicles, $\phi$, as a function of $\lambda$. A straightforward observation is that increased density drives $\phi$ to zero. Furthermore, we observe that $E[\phi(t)]$ is lower for settings with traffic lights (SL-A and GW-A), compared to no traffic lights (NL-A) setting. Vehicles held at junctions by traffic lights act as ``bridges'' for incident road segments. This is true even for low densities, with no vehicles approaching from other directions, as vehicles arriving at traffic lights still stop. In contrast, this is not the case for simple intersections and free-flow traffic. Note that for high densities the average proportion of isolated vehicles is higher for the SL-R scenario: congestion occurs in the central road segments and vehicles entering the network are more likely to be isolated.

\subsection{Critical Range}\label{sec:critical_range}

\begin{figure}[t]
\begin{center}
\includegraphics[width=0.8\linewidth]{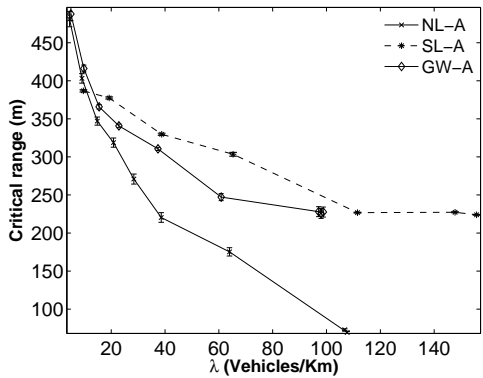}
\caption{Critical range vs. vehicle density.}
\label{fig:Critical_range_function_car_density}
\end{center}
\end{figure}

The critical range is shown in Fig.~\ref{fig:Critical_range_function_car_density} and it is, as one would expect, a decreasing function of vehicle density. We notice however that for the traffic lights scenario the critical range surprisingly flattens when we increase the vehicle density from 110 to 150 vehicles/Km. The critical range is the mean over the duration of each of the randomly seeded simulation runs. This implies that full connectivity is not guaranteed at all times. Recall that increasing the in-flow rate $f$ does not lead to an ever-increasing density. For example, for NL-A, the maximum average density achievable is approximately $110$ vehicles/Km, and the corresponding curve in Fig.~\ref{fig:Critical_range_function_car_density} is truncated accordingly.

%% To do
%% #17# We need a symbol for critical range

%% Too late...... :-(
%% #18# Once we have the look at the time-varying characteristics of connectivity,
%% we could perhaps argue that this is not a problem. We could promote to a paragraph here.

%% Too late.... :-(
%% #19# why not calculate the max critical range and the median and give side-by-side
%% those numbers?

With traffic lights, the vehicle density increases but the critical range does continue to decrease. Rather, there is a minimum value of a critical range of approximately 225 meters in the scenarios investigated here for SL and GW. The reason for this and the faster convergence of the critical range for GW is two-fold: (i) Central road density remains low and constant while in the peripheral roads it continues to increase, and (ii) gaps appear in connectivity (if the critical range were low) due to traffic lights. In other words, being interested in the critical range, we should consider the sparser part of the network, which is that of the central road segments. At the same time, an increase in the density in central roads does not necessarily result in lower critical range values: traffic lights essentially hold vehicles stopped, in longer queues, while the higher communication range values are needed to connect those stopped vehicles with the ones moving freely (not currently stopped by a traffic light). The designer of vehicular communication protocols should take into account this stop-and-go behavior, which is more pronounced in the SL scenario, as it causes inter-vehicle distance fluctuations and thus requires higher communication ranges to ensure connectivity.

%% ????????????
%% #21# It would be interesting to comment on what is the case for SL-R. If the same,
%% then let's just say that the observations are good for both cases.

%% Too late ... :-(
%% #22# If we cannot create a separate subsection, we should interleave
%% in the discussion evidence on how things change dynamically over time.
%% I.e., is the size (proportion) of the largest cluster changing significantly
%% over time?

\subsection{VC Market Penetration} \label{sec:market_penetration}

In this section, we study the impact of market penetration and the resultant background traffic, that is, the non-equipped vehicles, and their impact on connectivity. We model market penetration by declaring that each vehicle starting its trip in the simulation is equipped with vehicular communication hardware and software with probability $\rho$; we measure the connectivity level via simulations. As expected, the connectivity improves for increased market penetration: Increasing the proportion of equipped vehicles increases the probability to have an equipped  neighbor, and this way the overall connectivity level improves. However, increasing the background traffic does not only affect the proportion of equipped vehicles but impacts the overall traffic.
\begin{figure}[t]
\begin{minipage}{0.5\textwidth}
\centerline{\epsfig{figure=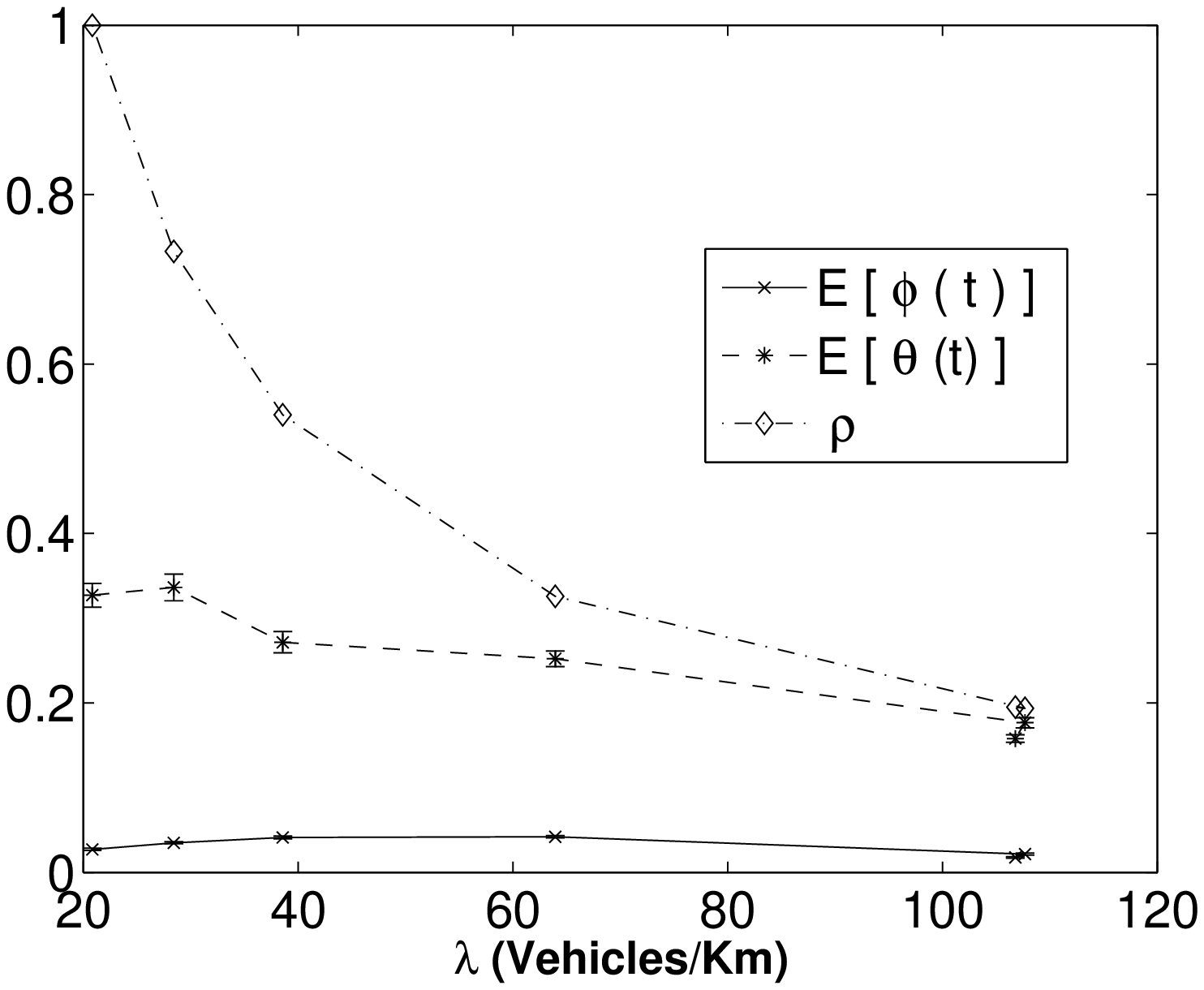,width=7cm}}
\caption{Scenario with no traffic lights and anti-diametric destinations.}
\label{fig:flow400_main_dest_no_tl}\end{minipage}
\begin{minipage}{0.5\textwidth}
\centerline{\epsfig{figure=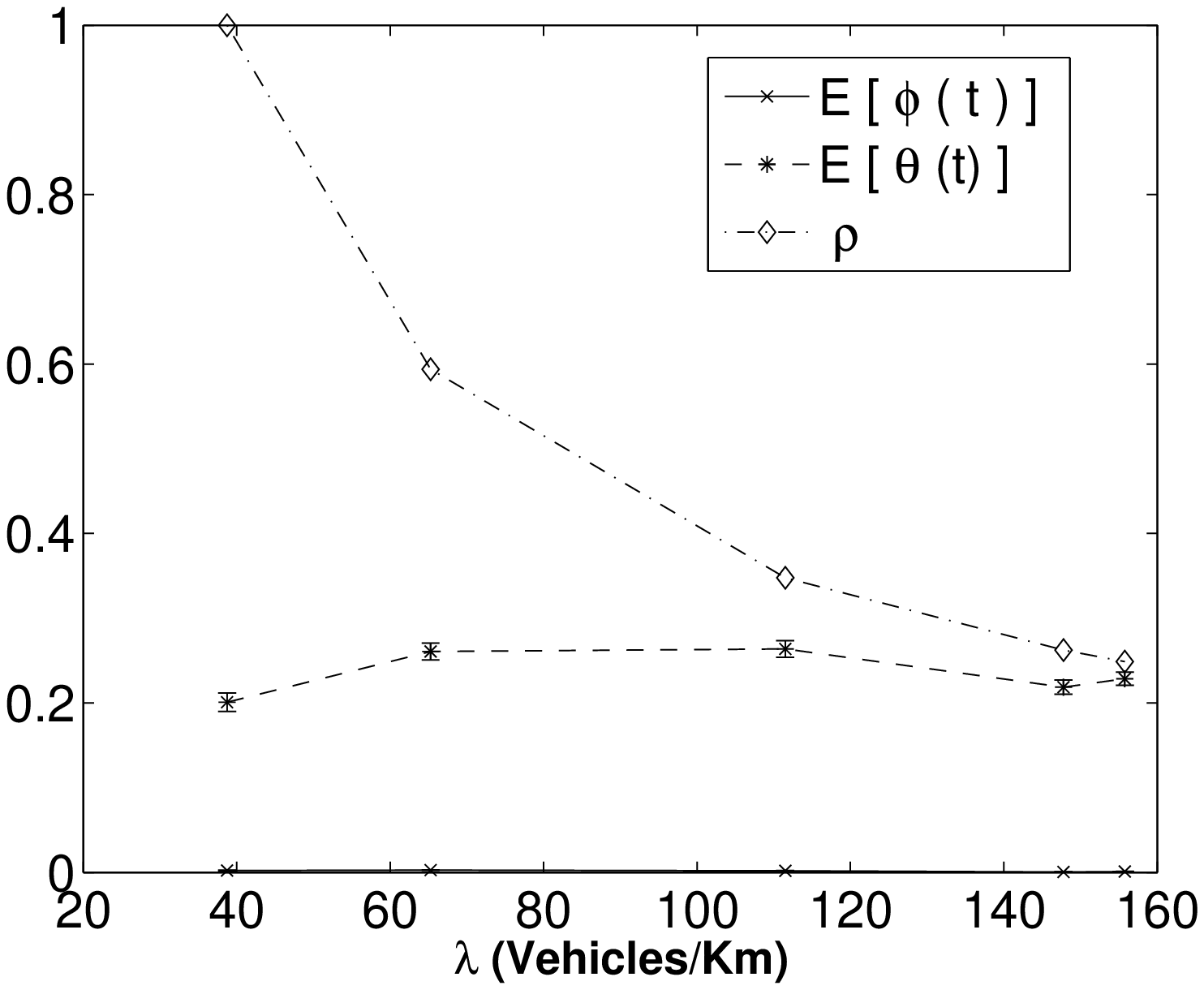,width=7cm}}
\caption{Scenario with traffic lights and anti-diametric destinations.}
\label{fig:flow400_main_dest_tl}\end{minipage}
\end{figure}

\begin{figure}[t]
\begin{minipage}{0.5\textwidth}
\centerline{\epsfig{figure=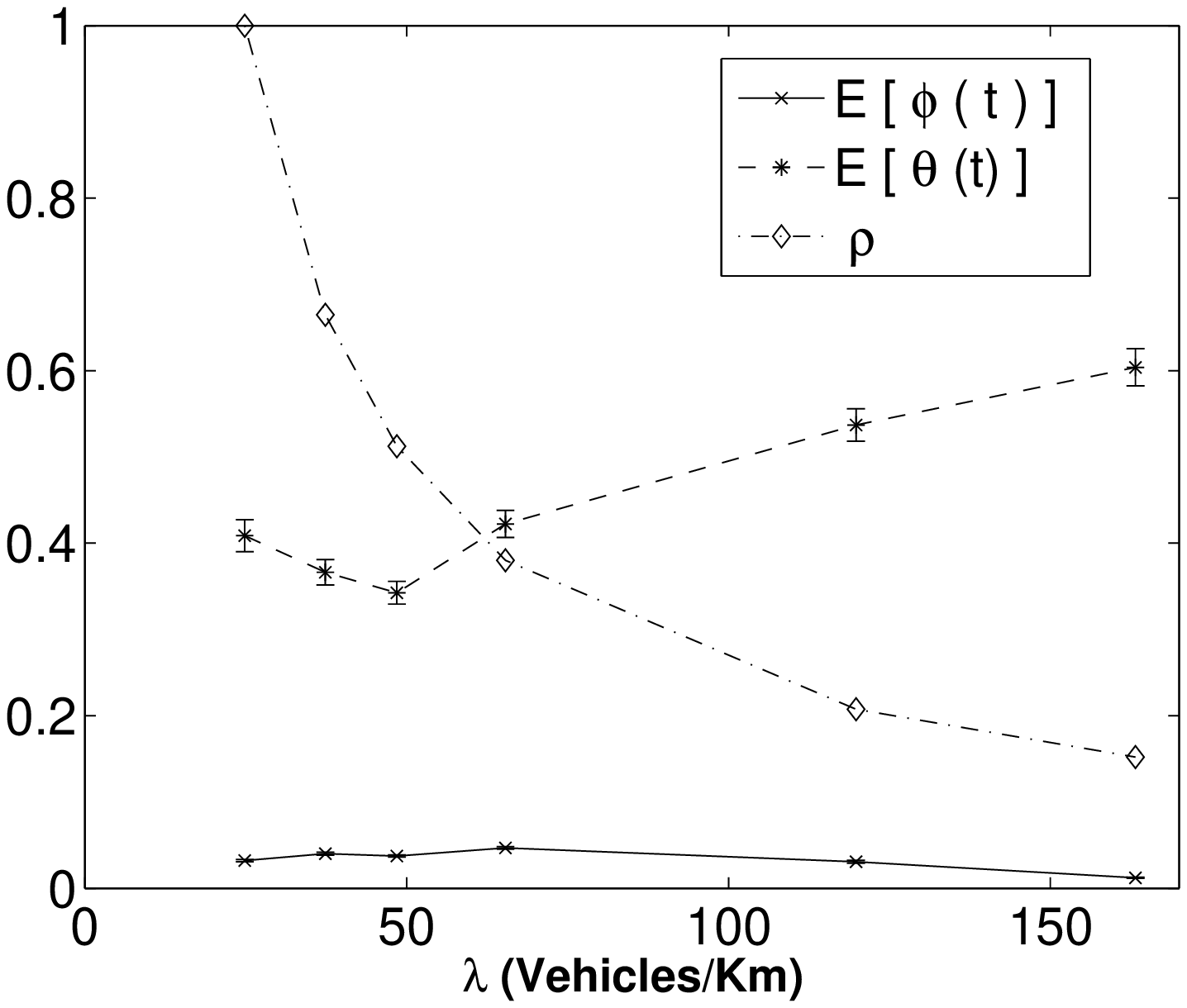,width=7cm}}
\caption{Scenario with no traffic lights and random destinations.}
\label{fig:flow400_dist_dest_no_tl}\end{minipage}
\begin{minipage}{0.5\textwidth}
\centerline{\epsfig{figure=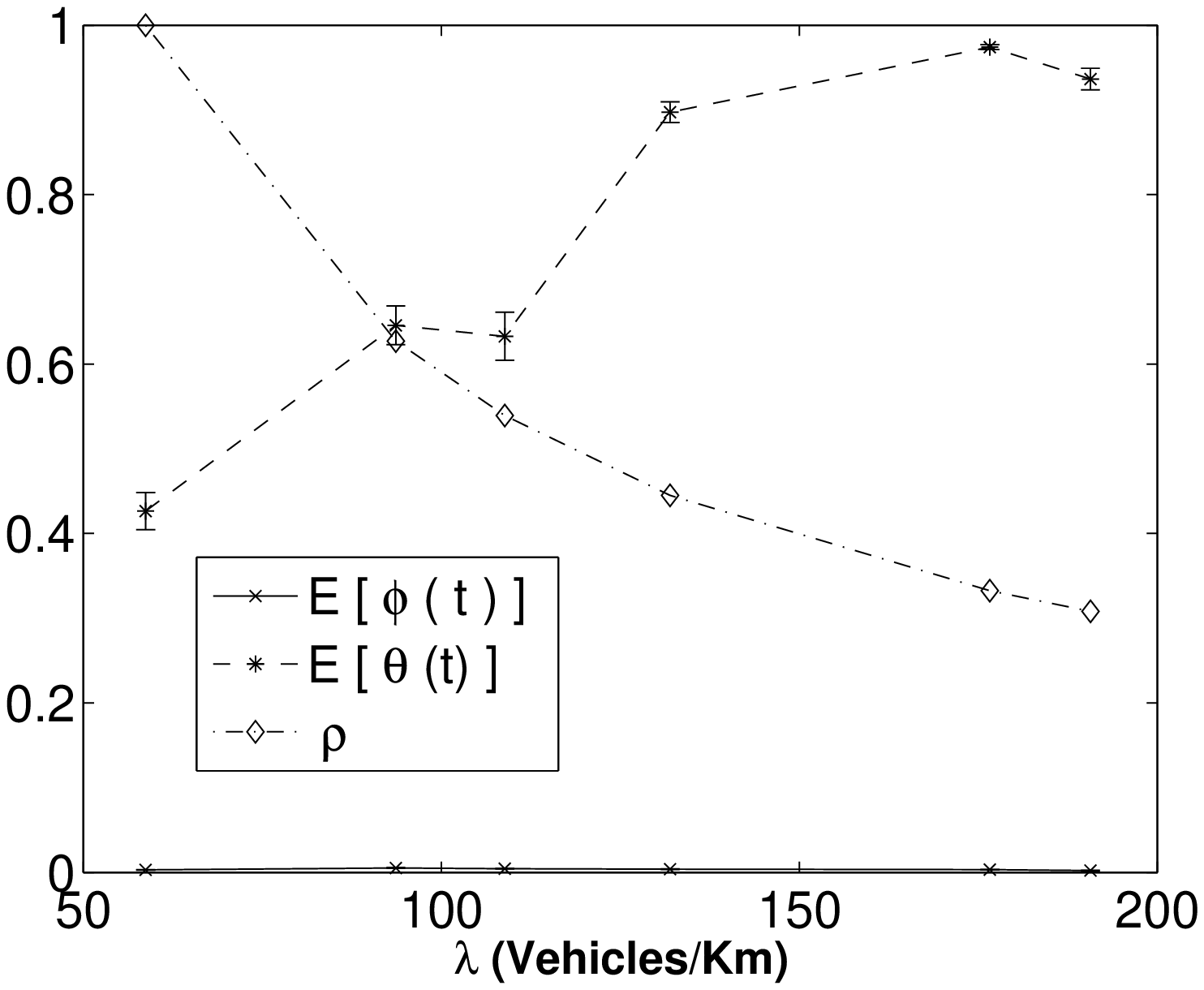,width=7cm}}
\caption{Scenario with synchronized traffic lights and random destinations.}
\label{fig:flow400_dist_dest_tl}\end{minipage}
\end{figure}

In Figs.~\ref{fig:flow400_main_dest_no_tl},~\ref{fig:flow400_main_dest_tl}, for scenarios with anti-diametric destinations: Background traffic has no significant effect on the connectivity level. Even though the average proportion of isolated vehicles tends to increase or remain constant,
the average proportion of vehicles in the biggest cluster increases somewhat, but the gain is not significant. In Figs.~\ref{fig:flow400_dist_dest_no_tl},~\ref{fig:flow400_dist_dest_tl}, for scenarios with random destinations: we obtain unexpected and quite interesting behavior of the average proportion of vehicles in the largest cluster. It increases significantly with background
traffic, especially in the case where we have traffic lights at intersections.

The question is how can the background traffic have an effect on the connectivity, and why we have this phenomenon only for the case when we have scenarios with distributed destinations. The key factor is the effect of vehicle trips and thus the distribution of destinations on the vehicle traffic and consequently their spatial distribution. Looking closely at the scenario with random destinations and no traffic lights (corresponding to the Fig.~\ref{fig:flow400_dist_dest_no_tl}), the proportion of vehicles in the biggest cluster decreases from ($\lambda = 24$  vehicles/Km, $\rho =
1$, $ E[\theta(t)] = 0.4$) to ($\lambda = 48$ vehicles/Km, $\rho = 0.51$, $E[\theta(t)] = 0.34 $) and increases to ($\lambda = 163$ vehicles/Km,
$\rho = 0.15$, $E[\theta(t)] = 0.6$). It appears that around the point corresponding to ($\lambda = 48$ vehicles/Km, $\rho = 0.51$, $E[\theta(t)]
= 0.34 $) a significant change takes place. We plot the spatial vehicle distribution over time around this point: We divide the maps in small zones and count the number of vehicles that went through in this zone over time.
\begin{figure}[t]
\begin{minipage}{0.5\textwidth}
\centerline{\epsfig{figure=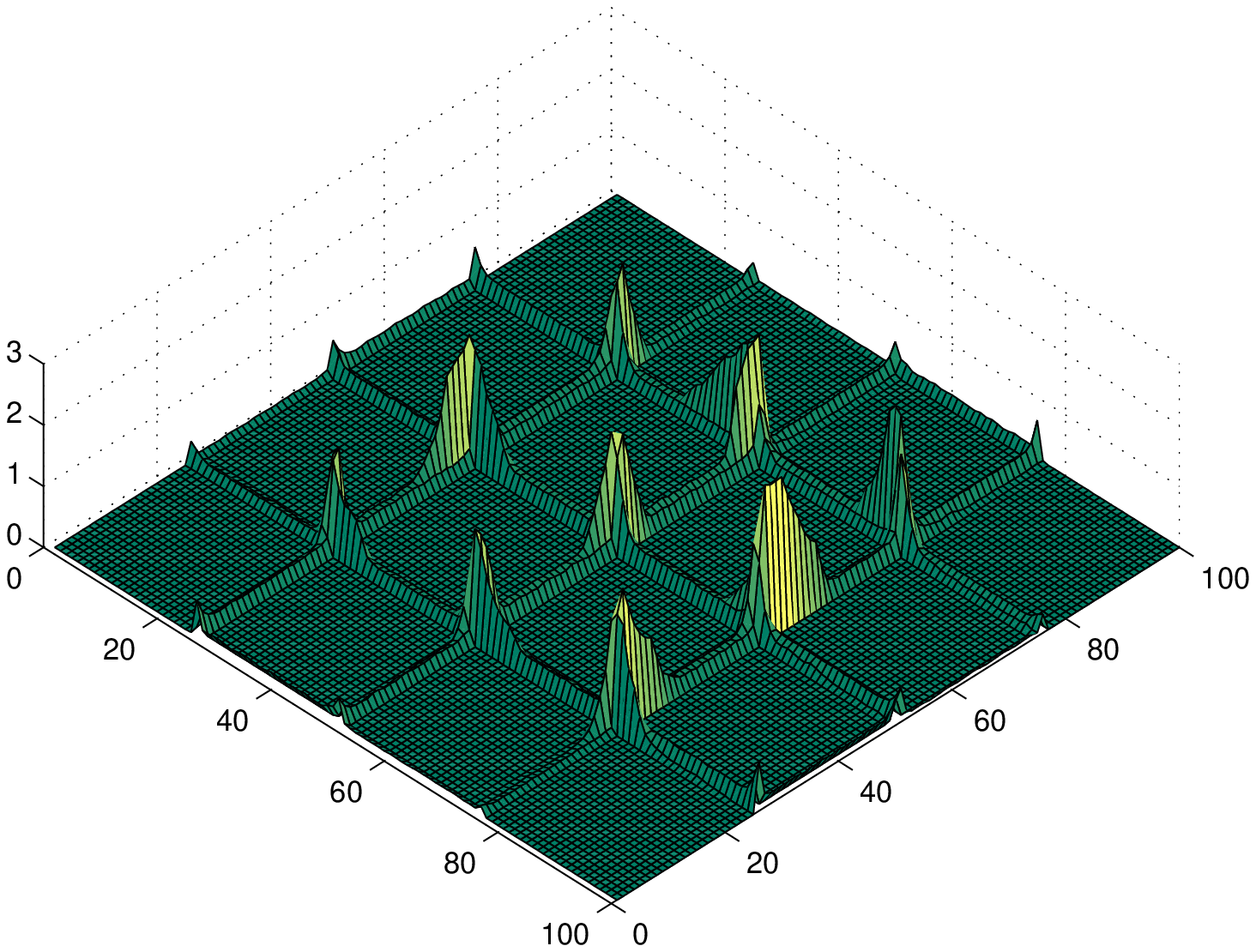,width=7cm}}
\caption{Vehicle distribution for $\lambda=37$ vehicles/Km and a NL-R scenario; the z-axis is the mean number of vehicles at each time step.}
\label{fig:Car_distribution_dist_dest_notl_flow500}\end{minipage}
\begin{minipage}{0.5\textwidth}
\centerline{\epsfig{figure=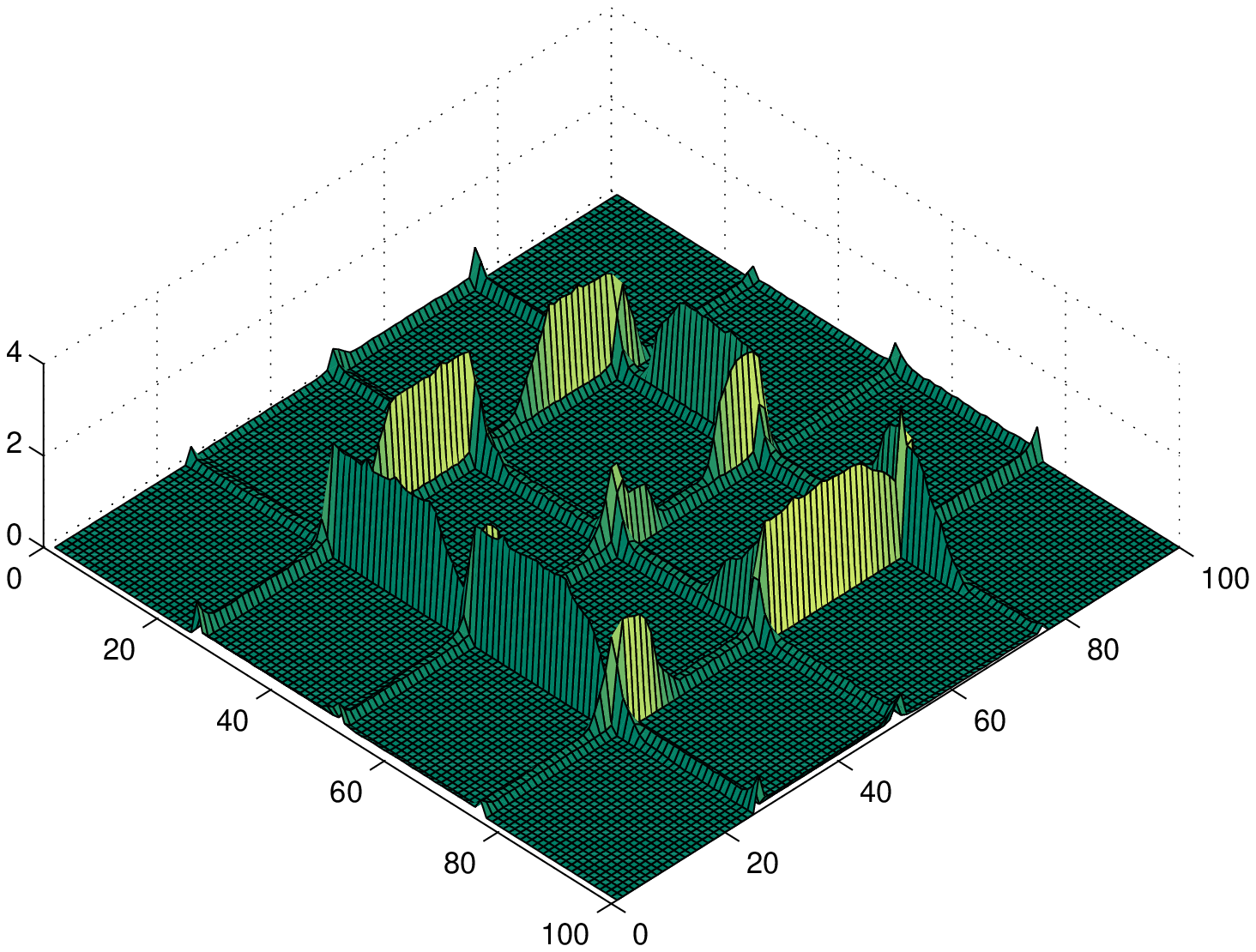,width=7cm}}
\caption{Vehicle distribution for $\lambda=65$ vehicles/Km and a NL-R scenario; the z-axis is the mean number of vehicles at each time step.}
\label{fig:Car_distribution_dist_dest_notl_flow600}\end{minipage}
\end{figure}

We see clearly that the spatial distribution of vehicles changes. In Fig.~\ref{fig:Car_distribution_dist_dest_notl_flow500}, there is significant concentration and increased vehicle density at intersections. While, in Fig.~\ref{fig:Car_distribution_dist_dest_notl_flow600}, the central
roads are congested and with an increased vehicle density. This is due to the fact that, in the distributed destinations scenarios, vehicles
make longer trips and are more likely to turn at intersections. This incurs a delay in the queues at intersections and leads to denser central roads (Fig.~\ref{fig:Car_density_Scenario4&5}). We see in Fig.~\ref{fig:Car_distribution_dist_dest_notl_flow600} that the vehicle traffic is essentially forming a square in the central zone. Therefore, equipped vehicles are more likely to be in range of each other and form large clusters, unlike the case of having vehicles mostly at intersections. Overall, the background traffic affects the mobility and distribution of equipped vehicles, it has an impact, sometimes positive, on the connectivity level and more precisely on the proportion of vehicles in the largest cluster.

\section{Connectivity with fixed Road Side Units}
\label{Sec:ConnectivityRSU}

We investigate the effects of \textit{road side units} (RSUs) on connectivity of VANETs. We quantify improvements in connectivity due to RSUs using the  measures defined in the previous sections, considering NL-A and SL-A scenarios, in order to compare the connectivity levels between RSU-assisted configurations and without RSUs present in the network. We do not dwell on the exact form of communication via RSUs, as this will depend on the applications that will be deployed in specific systems. For the purpose of this investigation, it is assumed that all RSUs are connected over wired or other fixed communication links, e.g., the Internet. Hence, any two vehicles out of range of each other but within range still communicate as long as they both have an RSU in range. Again, how this will be done depends on future applications; as one example, a vehicle can pass a hazardous traffic alert to an RSU (possibly over multiple hops, i.e., vehicles), and the road side infrastructure can relay this alert via other RSUs to vehicles that are otherwise disconnected. The RSUs are placed at intersections and have a distance of 400m between each other. We assume that they have the same communication range as the vehicles. The placement of RSUs at the intersections is motivated by the already existing infrastructure (traffic lights, electricity, etc.) at these locations and has not been optimized with respect to connectivity.

\subsection{Impact of RSUs on the biggest cluster}

\begin{figure}[t]
\begin{minipage}{0.5\textwidth}
\centerline{\epsfig{figure=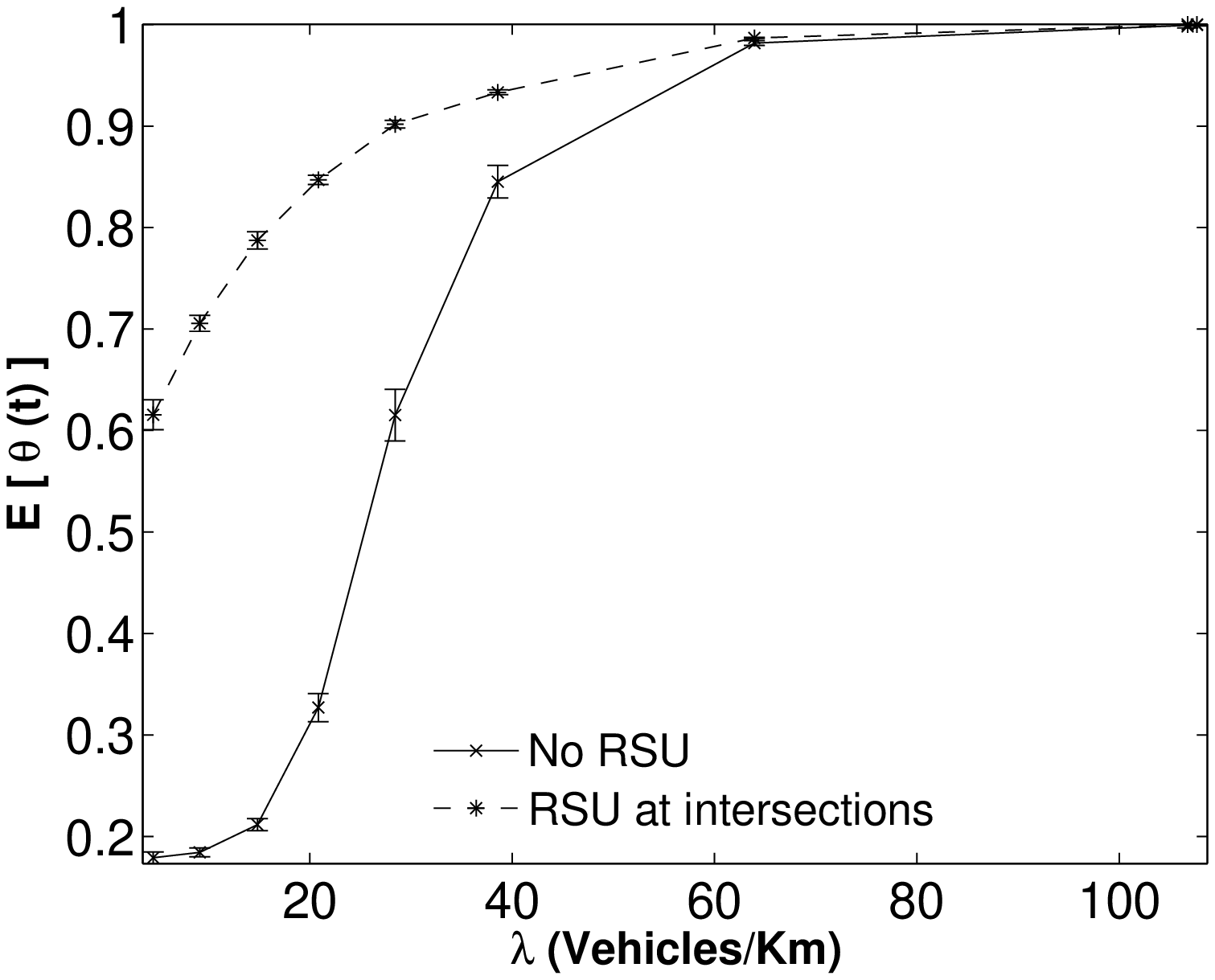,width=7cm}}
\caption{Proportion of vehicles in the biggest cluster vs. vehicle density in scenario NL-A.}
\label{fig:Scenario1_Biggest_cluster_rsu}\end{minipage}
\begin{minipage}{0.5\textwidth}
\centerline{\epsfig{figure=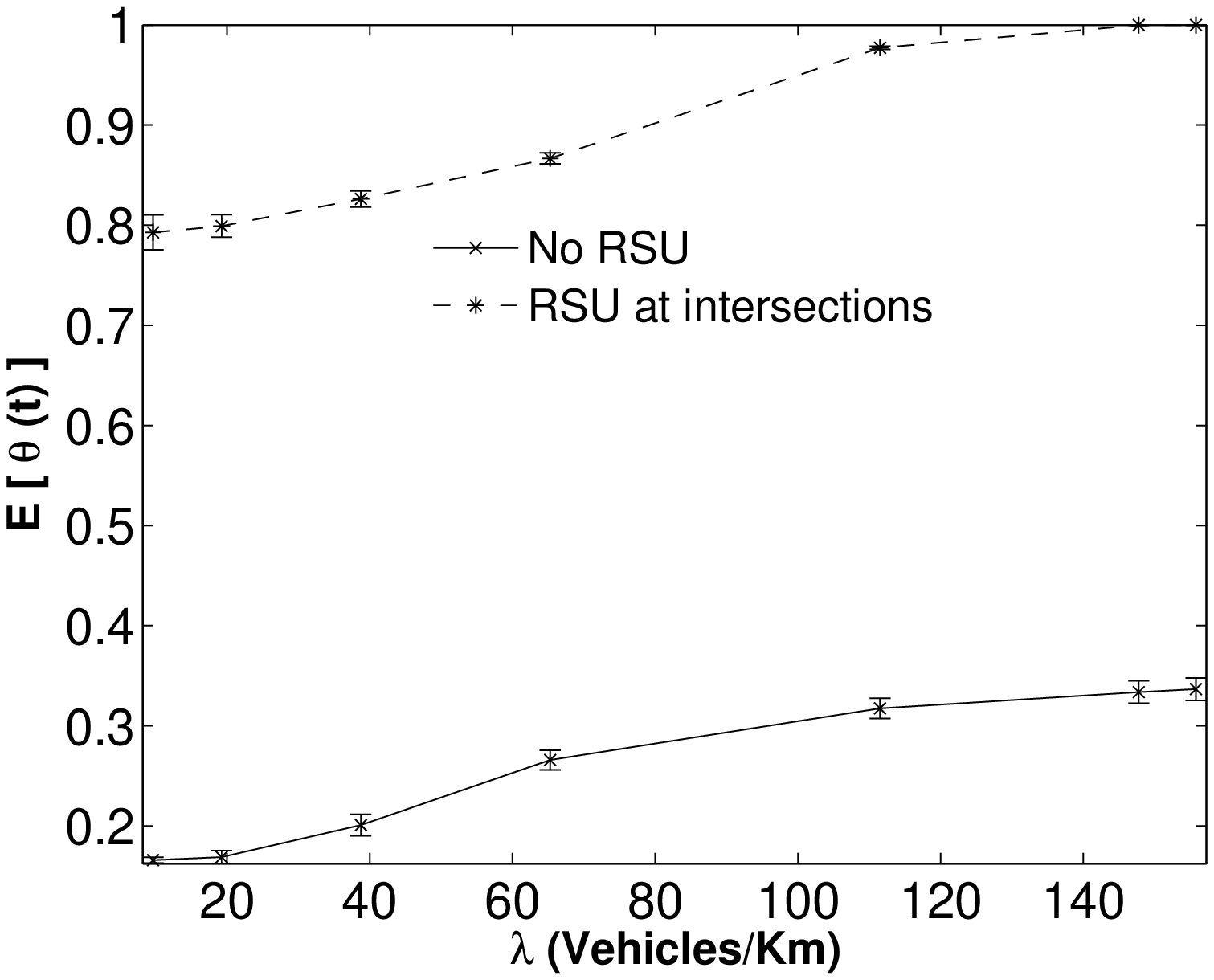,width=7cm}}
\caption{Proportion of vehicles in the biggest cluster vs. vehicle density in scenario SL-A.}
\label{fig:Scenario2_Biggest_cluster_rsu}
\end{minipage}
\end{figure}

We observed in Sec.~\ref{Sec:EvalConneAd-hocNet} that the network percolates for the scenarios with simple intersections while it does not do so when
there are traffic lights. We naturally expect to see a more significant connectivity gain when introducing RSUs to the scenario with traffic lights.
The results in Fig.~\ref{fig:Scenario1_Biggest_cluster_rsu} and \ref{fig:Scenario2_Biggest_cluster_rsu} confirm our expectations. When we
consider Scenario NL-A at the subcritical phase, we notice that there is a significant improvement in connectivity. At the supercritical phase, the
connectivity improves slightly just above the critical density. However, the difference vanishes at higher densities.

The difference in connectivity gain between the two phases can be explained as follows. In the subcritical phase, the vehicles form small clusters disconnected from each other; i.e., vehicles belonging to the same cluster can communicate but they are isolated from other vehicles. Nonetheless, a vehicle connected to a RSU essentially acts a relay for all other vehicles in the cluster and links its peers to other clusters; this forms a big cluster consisting of small interconnected ones and the proportion of vehicles in the largest cluster increases. In the super-critical phase, a percolating cluster, containing the majority of vehicles communicate in a pure ad-hoc way, already exists. This percolating cluster is almost surely connected with an RSU. Thus, the slight connectivity improvement results from the isolated small clusters which are connected to an RSU and can then join the main percolating component. In the SL-A scenarios, the proportion of vehicles in the biggest cluster increases drastically with the assistance of RSUs: the network never percolates in a pure ad hoc (vehicle to vehicle) manner but the RSUs link the isolated clusters to form a percolating one.

\subsection{Impact of RSUs on the isolated vehicles}

\begin{figure}[t]
\begin{center}
\includegraphics[width=0.8\linewidth]{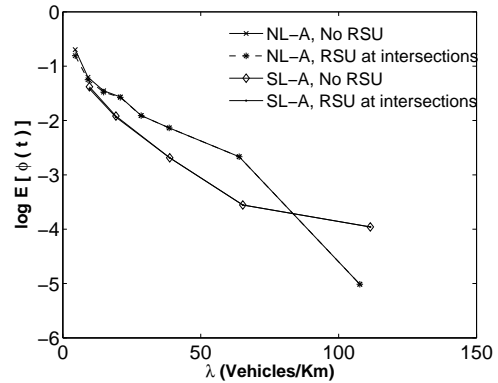}
\caption{Proportion of isolated vehicles vs. vehicle density in both NL-A and SL-A.}
\label{fig:Scenario1&2_isolatedcars_rsu}
\end{center}
\end{figure}

Fig.~\ref{fig:Scenario1&2_isolatedcars_rsu} shows the fraction of isolated vehicles as a function vehicle density in pure and hybrid (with RSUs) ad-hoc networks. We notice that the proportion of isolated vehicles does not change significantly when introducing RSUs at intersections. Knowing that the intersections are the most congested spots in the domain, it follows that the isolated vehicles are more likely to be in the middle of the road or at the entering points of the domain. Therefore, the RSUs placed at intersections do not benefit these isolated vehicles. For improved connectivity in such cases a different RSU placement strategy can be used; for example, one placing RSUs at the middle of road segments.

\section{Conclusion} \label{Sec:conclusion}

In this paper, we provide a framework for the study of vehicular connectivity in urban scenarios, and we show the use of percolation theory for that purpose. We provide an extensive set of simulations that reveal the impact of main vehicle and transportation parameters and factors, such as vehicle density, traffic light, ``background'' vehicle traffic (in the case of low market penetration rate for vehicular communications), and RSUs that can facilitate vehicle connectivity. These investigations result in a range of interesting findings, among which we summarize here the basic ones.
\begin{enumerate}
\item We can infer the global connectivity level based on the local vehicle density, under certain conditions (near-uniform distribution of vehicle traffic). We confirm through simulations the existence of a critical density (shown by percolation theory), above which the connectivity significantly improves.
\item A well-connected vehicle-to-vehicle network can be formed at all times, even with relatively sparsely placed vehicles. In fact, good (or even full) vehicular connectivity is possible without road congestion (due to high vehicle density) and relatively low communication range, for example, 25\% of a road segment length.
\item Traffic lights have an important impact on connectivity. In particular, on the one hand, the commonly observed accumulation of vehicles at red traffic lights can be beneficial for connectivity as it can create vehicle-to-vehicle meeting points. On the other hand, this clustering has the drawback of increasing the distance between equipped vehicles and its fluctuations. Thus, the largest cluster size remains low.
\item For low market penetration, vehicular communication connectivity can be significantly different from that encountered when all vehicles are equipped with communicating radios. This is especially due to the role of unequipped vehicles acting as ``background'' traffic, that is physically occupying space and altering the spatial distribution, mobility, and finally connectivity of the equipped vehicles.
\item RSUs do \emph{not} significantly improve connectivity in all scenarios. On the one hand, RSUs at intersections do not reduce the proportion of isolated vehicles. On the other hand, if the vehicular communication network is at its super-critical phase (but not fully connected yet), RSUs do not significantly increase the connectivity (i.e., the size of the largest cluster).
\end{enumerate}

These results show that connectivity cannot be taken for granted. In fact, it should be carefully assessed before the deployment of any real-world
application. The approach provided by this paper shows how to do so. This is a significant first step, bringing in microscopic simulation and theoretical results. In terms of future work, we intend to integrate a sophisticated radio model, the use of directional antennas, and the influence of the MAC layer. At the same time, we will produce additional theoretical approximations, on aspects such as the critical range, or modeling the traffic
light regimes, for example, calculating the connectivity probability for those scenarios, as well as those with RSUs.

\bibliography{autonet08}
\bibliographystyle{plain}
\end{document}